\documentstyle[eqsecnum,aps,preprint,psfig]{revtex}
\tightenlines

\def\be{\begin{eqnarray}}
\def\ee{\end{eqnarray}}

\def\bb{\bbox}

\begin{document}
\draft
 \title  {\bf Continuum quasiparticle random phase approximation and the
   time-dependent Hartree-Fock-Bogoliubov approach
}

\author{E. Khan$^{a)}$,  N. Sandulescu$^{b),c),d)}$,  M.
Grasso$^{a)}$, Nguyen Van Giai$^{a)}$}

\vspace {03mm}

\address{
{\it a) Institut de Physique Nucl\'eaire, IN$_{2}$P$_{3}$-CNRS, 91406 Orsay,
France}\\
{\it b) Institute for Physics and Nuclear Engineering, P.O. Box MG-6, 76900
Bucharest, Romania}\\
{\it c) Research Center for Nuclear Physics, Osaka University, 567-0047 
Osaka, Japan}\\
{\it d) Royal Institute of Technology, SCFAB, SE-10691, Stockholm, Sweden}
}


\maketitle

\begin{abstract}
Quadrupole excitations of neutron-rich nuclei are analyzed by using the
linear response method in the Quasiparticle Random Phase Approximation
(QRPA). The QRPA response is derived starting from the time-dependent
Hartree-Fock-Bogoliubov (HFB) equations. The residual interaction between
the quasiparticles is determined consistently from the two-body force
used in the HFB equations, and the continuum coupling is treated exactly. 
Calculations are done for the neutron-rich oxygen isotopes. 
It is found that pairing correlations affect the low-lying states, and that
a full treatment of the continuum can change the structure of the states in
the giant resonance region.

\end{abstract}

\vskip 0.5cm


\section{Introduction}

 The collective excitations of atomic nuclei in the presence of pairing
 correlations is usually described in the quasiparticle-Random Phase
 Approximation (QRPA) \cite{rs80}. Although the QRPA was applied to nuclear
 physics more than 40 years ago \cite{av60,km60,ba60}, recently there is a
 renewed interest on its grounds, generated mainly by the studies of
 unstable nuclei close to the drip line. In these nuclei characterized by a
 small nucleon separation energy, the excited states are strongly influenced
 by the coupling with the quasiparticle (qp) continuum configurations. Among
 the configurations of particular interest are the two-qp states in which
 one or both quasiparticles are in the continuum. In order to describe such
 excited states within QRPA one needs a proper treatment of the continuum
 coupling, which is missing in the usual QRPA calculations based on a
 discrete qp spectrum.

  In nuclei close to the drip lines one expects also a strong connection
  between the excitations of the system and the properties of the ground 
  state, which may present such specificities as neutron skins. 
 Therefore, in addition to the qp 
 spectrum, the residual interaction used in QRPA should
 be determined  from the same two-body force as it is done in the 
 self-consistent continuum RPA calculations \cite{be75,lg76,gi81}.
 
 In the past years several attempts \cite{ka98,ha82,ma01} 
 have been made to describe 
 consistently both the pairing correlations and the continuum 
 coupling within QRPA. Thus, in 
 Ref.\cite{ha82} 
 a QRPA approach was recently developed in which the effect of
 the continuum is calculated exactly for the particle-hole
 excitations whereas in the particle-particle channel the
 active space is limited to the bound states close to the
 Fermi level. 

 A continuum qp linear response approach in which 
 the continuum is included also in the particle-particle channel  
 was studied in Ref.\cite{ma01}, but in the calculations
 the ground state mean field is fixed independently of the 
 residual interaction.

 In this paper we present the first continuum QRPA calculations with 
 the single-particle spectrum and the residual interaction determined 
 from the same effective two-body force. The ground state is calculated 
 using the continuum HFB approach \cite{gr01} with  the mean field and
 the pairing field described by a Skyrme interaction and a density dependent 
 delta force, respectively. Based on the same HFB energy functional we
 derive the QRPA response function in coordinate space. 
 The QRPA response is constructed by using real
 energy solutions for the continuum HFB spectrum. The calculations are
 done for the neutron-rich oxygen isotopes. 

 In Section II we present the continuum QRPA formalism, we specialize the
corresponding equations to systems with spherical symmetry and we 
discuss the energy-weighted sum rule in QRPA. The
application of the present theory to neutron-rich oxygen isotopes is done 
in Section III. Section IV contains the concluding remarks.

\section{Formalism}

\subsection{Derivation of the generalized Bethe-Salpeter equation}
The coordinate space formalism is naturally adapted to treat properly the
coupling to the continuum states. In this section we derive the QRPA
equations in coordinate space as the small amplitude limit of the perturbed
time-dependent HFB equations. 
We start from the time-dependent HFB (TDHFB) equations \cite{rs80}:

\begin{equation}\label{eq:tdhfb}
i\hbar\frac{\partial{\cal R}}{\partial t}=[{\cal H}(t) + 
{\cal F}(t),{\cal R}(t)]
\end{equation}
where ${\cal R}$ and ${\cal H}$ are the time-dependent generalized
density and HFB hamiltonian. 
The external periodic field ${\cal F}$ is given by
\begin{equation}\label{eq:pert}
{\cal F} = F e^{-i\omega t} + h.c.
\end{equation}
where $F$ includes both particle-hole and two-particle transfer
operators:
\begin{equation}\label{eq:extpart}
F=\sum_{ij} F^{11}_{ij} c_{i}^{\dagger}c_{j}+\sum_{ij}
(F^{12}_{ij} c_{i}^{\dagger}c_{j}^{\dagger}+ F^{21}_{ij} c_{i}c_{j})
\end{equation}
and $c_{i}^{\dagger}$, $c_{i}$ are the particle creation and annihilation
operators, respectively.
Assuming that the external field induces small oscillations around the
stationary solution of the HFB equations, 
\begin{equation}\label{eq:pertr}
        {\cal R}(t) = {\cal R}^0+ {\cal R}' e^{-i\omega t} + h.c.
\end{equation}

\begin{equation}\label{eq:perth}
        {\cal H}(t) = {\cal H}^0+ {\cal H}' e^{-i\omega t} + h.c.
\end{equation}
the TDHFB equation (\ref{eq:tdhfb}) becomes
\begin{equation}\label{eq:lin}
  	\hbar\omega{\cal R}'=[{\cal H}',{\cal R}^0] + [{\cal H}^0,{\cal
	R}']+[F,{\cal R}^0]
\end{equation}
The generalized density variation has the form:
\begin{equation}\label{eq:denspart}
	{\cal R}'_{ij}=\left(
	\begin{array}{cc}
	\rho'_{ij} & \kappa'_{ij} \\
	\bar{\kappa}'_{ij} & -\rho'_{ji} \\
	\end{array}
	\right)
\end{equation}
where $\rho'_{ij} = \left<0|c^{\dagger}_jc_i|'\right>$
is the variation of the particle density,
$\kappa'_{ij} =\left<0|c_jc_i|'\right>$ and $\bar{\kappa}'_{ij} =
\left<0|c^{\dagger}_jc^{\dagger}_i|'\right>$ are the
fluctuations of the pairing tensor associated to the pairing vibrations and
$\mid ' \rangle$ denotes the change of the ground state wavefunction
$|0>$ due to the external field. Instead of the variation of one quantity
in RPA ($\rho'$) , we now have to know the variations
of three
independent quantities in QRPA, namely $\rho'$, $\kappa'$ and $\bar{\kappa}'$.

It is convenient to solve Eq. (\ref{eq:lin}) in the qp representation in
which 
both ${\cal H}^0$ and ${\cal R}^0$ are diagonal\cite{rs80}. We have now to
express all quantities of Eq. (\ref{eq:lin}) in this representation.   
The matrix ${\cal R}'$ becomes 
off-diagonal because of the TDHFB condition  
${{\cal R}'}^2={\cal R}'$ imposed on Eq.(\ref{eq:pertr}):
\begin{equation}\label{eq:rprime}
	\tilde{{\cal R}}'_{ij}=\left(
	\begin{array}{cc}
	0 & \tilde{{\cal R}}'^{12}_{ij} \\
	\tilde{{\cal R}}'^{21}_{ij}& 0 \\
	\end{array}\right) = 
	\left(\begin{array}{cc}
	0 &  \left<0|\beta_j\beta_i|'\right> \\
	\left<0|\beta^{\dagger}_j\beta^{\dagger}_i|'\right> &0 \\
	\end{array}\right)
\end{equation}
where $\beta^{\dagger}_i$, $\beta_i$ are respectively the qp creation and
annihilation operators of an HFB state $i$ with energy $E_i$.
Consequently, Eq. (\ref{eq:lin}) gives
\begin{equation}\label{eq:r12}
	\tilde{{\cal R}}'^{12}_{ij}=\frac{\tilde{{\cal
	H}}'^{12}_{ij}+\tilde{F}^{12}_{ij}}
	{\hbar\omega-(E_i+E_j)},       
\end{equation}

\begin{equation}\label{eq:r21}
	\tilde{{\cal R}}'^{21}_{ij}=-\frac{\tilde{{\cal
	H}}'^{21}_{ij}+\tilde{F}^{21}_{ij}}
	{\hbar\omega+(E_i+E_j)}.       
\end{equation}
Here, 
$\tilde{\cal H}'$ and 
$\tilde{ F}$ stand for  
${\cal H}'$ and ${ F}'$ in the qp representation.

We now proceed to calculate $\tilde{{\cal H}}'^{12}_{ij}$ and
$\tilde{{\cal H}}'^{21}_{ij}$.
The variations of the particle and
pairing densities 
in coordinate representation are defined by:

\begin{equation}\label{eq:rhor}
\rho'\left({\bf r}\sigma\right) = 
	\left<0|\psi^{\dagger}\left({\bf r}\sigma\right)
	\psi\left({\bf r}\sigma\right)|'\right>
\end{equation}

\begin{equation}\label{eq:rhotildr}
        \kappa'\left({\bf r}\sigma\right) = 
	\left<0|\psi\left({\bf r}\bar{\sigma}\right)
	\psi\left({\bf r}\sigma\right)|'\right>
\end{equation}

\begin{equation}\label{eq:rhotildbarr}
        \bar{\kappa}'\left({\bf r}\sigma\right) =
	\left<0|\psi^{\dagger}\left({\bf r}\sigma\right)
	\psi^{\dagger}\left({\bf r}\bar{\sigma}\right)|'\right>
\end{equation}
where
$\psi^{\dagger}\left({\bf r}\sigma\right)$ is
the particle creation operator in coordinate space and
$\psi^{\dagger}\left({\bf r}\bar{\sigma}\right)$=
$-2\sigma \psi^{\dagger}\left({\bf r}-\sigma\right)$ is its time reversed
counterpart.
The relation between the $\psi^{\dagger}, \psi$ and $\beta^{\dagger}, \beta$
operators is: 
\begin{equation}\label{eq:psi}
\psi^{\dagger}\left({\bf r}\sigma\right)=
\sum_{k} U_{k}\left({\bf r}\sigma\right)\beta_k+
V^{*}_{k}\left({\bf r}\sigma\right)\beta^{\dagger}_k
\end{equation}
where $U_k$ and $V_k$ are the two components of the HFB wave
function of the qp state with energy $E_k$.

Introducing Eq.(\ref{eq:psi}) and its hermitian conjugate into 
Eqs.(\ref{eq:rhor}-\ref{eq:rhotildbarr}) one gets with the help of 
Eq. (\ref{eq:rprime}) :

\begin{equation}\label{eq:rhopertr}
{\rho'}_\alpha\left({\bf r}\sigma\right)=\sum_{ij}
{\cal U}^{\alpha1}_{ij}\left({\bf r}\sigma\right)\tilde{\cal R}'^{12}_{ij}+
{\cal U}^{\alpha2}_{ij}\left({\bf r}\sigma\right)\tilde{\cal
  R}'^{21}_{ij},~~~\alpha=1,2,3
\end{equation}   

where we have introduced the following notation for 
the density variations:

\begin{equation}\label{eq:rhodef}
{\bb{\rho}'}=\left(
        \begin{array}{c}
         \rho'_1 \\
         \rho'_2 \\
         \rho'_3 \\
        \end{array}
        \right)
	=\left(
        \begin{array}{c}
         \rho' \\
         \kappa' \\
         \bar{\kappa}' \\
        \end{array}
        \right)
\end{equation}

and the 3 by 2 matrices ${\cal U}_{ij}$ are defined by: 
\begin{equation}\label{eq:ubig}
{\cal U}_{ij}({\bf r}\sigma)=\left(
        \begin{array}{cc}
 U_i({\bf r}\sigma)V_j({\bf r}\sigma)~~ & 
 U^*_j({\bf r}\sigma)V^*_i({\bf r}\sigma)
\\
 U_i({\bf r}\sigma)U_j({\bf r}\bar{\sigma})~~ & 
 V^*_i({\bf r}\sigma)V^*_j({\bf r}\bar{\sigma})
\\ 
-V_i({\bf r}\sigma)V_j({\bf r}\bar{\sigma})~~ &
-U^*_i({\bf r}\sigma)U^*_j({\bf r}\bar{\sigma})
\\
\end{array}
\right)
\end{equation}
Here, we have used the same notation as introduced before for 
the time reversed particle operators, 
i.e., $f({\bf r}\bar{\sigma})$=$-2\sigma f({\bf r}-\sigma)$.

Next, we must calculate the variation $\tilde{{\cal H}}'$ of the HFB
hamiltonian in the qp representation. 
This is obtained from the corresponding quantity in coordinate
representation through the transformation \cite{rs80}:
\begin{equation}\label{eq:hqpart}
        \tilde{{\cal H}}'= W^\dagger{\cal H}'W
\end{equation}
where the matrix $W$ is defined by:
\begin{equation}\label{eq:w}
        W=\left(
        \begin{array}{cc}
        U  & V^* \\
        V & U^* \\
        \end{array}
        \right)
\end{equation}
One thus gets:

\begin{equation}\label{eq:hqparthco12}
\tilde{{\cal H}}'^{12}_{ij}=\int d{\bf r}\sum_{\sigma}
\bar{{\cal U}}^{*11}_{ij}({\bf r}\sigma){\cal H}'^{11}({\bf r})-
\bar{{\cal U}}^{\dagger21}_{ij}({\bf r}\sigma){\cal H}'^{12}({\bf r})-
\bar{{\cal U}}^{\dagger31}_{ij}({\bf r}\sigma){\cal H}'^{21}({\bf r})
\end{equation}

\begin{equation}\label{eq:hqparthco21}
\tilde{{\cal H}}'^{21}_{ij}=\int d{\bf r}\sum_{\sigma}
\bar{{\cal U}}^{*12}_{ij}({\bf r}\sigma){\cal H}'^{11}({\bf r})-
\bar{{\cal U}}^{\dagger22}_{ij}({\bf r}\sigma){\cal H}'^{12}({\bf r})-
\bar{{\cal U}}^{\dagger32}_{ij}({\bf r}\sigma){\cal H}'^{21}({\bf r})
\end{equation} 
where $\bar{{\cal U}}_{ij}={\cal U}_{ij}-{\cal U}_{ji}$.
This antisymmetric combination appears by taking into account
the relation ${\cal H}'^{22}_{ij}$=-${\cal H}'^{11}_{ji}$ \cite{rs80}.
Similar equations hold for the matrix elements of the external
field (\ref{eq:extpart}).

In coordinate representation the variation of the HFB hamiltonian 
is expressed in terms of the second derivatives of the HFB
energy functional ${\cal E}$[$\rho,\kappa,\bar{\kappa}$]
with respect to the densities. Thus, in our matrix notation we 
can write (see appendix) :
\begin{equation}\label{eq:hvar}
\bb{{\cal H}}'=	\left(\begin{array}{c}
	 {\cal H}'^{11} \\
	 {\cal H}'^{12}\\
	 {\cal H}'^{21}   \\
	\end{array}
	\right)=\bf{V}\bb{\rho}'
\end{equation}
where $\bf{V}$ is the residual interaction matrix, namely :
\begin{equation}\label{eq:vres}
{\bf{V}}^{\alpha\beta}({\bf r}\sigma,{\bf r}'{\sigma}')=
\frac{\partial^2{\cal E}}{\partial{\bf{\rho}}_\beta({\bf r}'{\sigma}')
\partial{\bf{\rho}}_{\bar{\alpha}}({\bf r}\sigma)},~~~\alpha,\beta = 1,2,3.
\end{equation}
Here, the notation $\bar{\alpha}$ means that whenever $\alpha$ is 2 or 3
then $\bar{\alpha}$ is 3 or 2. 

It should be noted that in the three dimensional space, the first
dimension represents the particle-hole (ph) subspace, the second the
particle-particle (pp) one, and the third the hole-hole (hh) one. This is
due to the definitions (\ref{eq:rhor}-\ref{eq:rhotildbarr}).

Using Eqs. (\ref{eq:r12}), (\ref{eq:r21}), (\ref{eq:rhopertr}) and 
(\ref{eq:hqparthco12}-\ref{eq:hvar}), we
finally land on the coupled equations :

\begin{equation}\label{eq:rhoeq}
\bb{\rho}'=\bf{G}_0\bf{V}\bb{\rho}'+\bf{G}_0\bf{F}
\end{equation}

where $\bf{F}$ is the three dimensional column vector 

\begin{equation}\label{eq:f}
\bf{{F}}=	\left(\begin{array}{c}
	 {F}^{11} \\
	 {F}^{12}\\
	 {F}^{21}   \\
	\end{array}
	\right)
\end{equation}

and $\bf{G}_0$ is the unperturbed Green function defined by:

\begin{equation}\label{eq:g0}
{\bf{G}_0}^{\alpha\beta}({\bf r}\sigma,{\bf r}'{\sigma}';\omega)=
\sum_{ij} \frac{{\cal U}^{\alpha 1}_{ij}({\bf r}\sigma)
\bar{{\cal U}}^{*\beta 1}_{ij}({\bf r}'\sigma')}{\hbar\omega-(E_i+E_j)+i\eta}
-\frac{{\cal U}^{\alpha 2}_{ij}({\bf r}\sigma)
\bar{{\cal U}}^{*\beta 2}_{ij}({\bf r}'\sigma')}{\hbar\omega+(E_i+E_j)+i\eta}
\end{equation}

By definition the QRPA Green function $\bf{G}$ relates the perturbing
external field to the density change: 
\begin{equation}\label{eq:g}
\bb{\rho}'=\bf{G}\bf{F}~.
\end{equation}
Combining with Eq.(\ref{eq:rhoeq}) we obtain 
the generalized Bethe-Salpeter equation :
\begin{equation}\label{eq:bs}
\bf{G}=\left(1-\bf{G}_0\bf{V}\right)^{-1}\bf{G}_0=\bf{G}_0+\bf{G}_0\bf{V}\bf{G}
\end{equation}

In the case of transitions from the ground state to excited states within
the same nucleus, only the (ph,ph) component of $\bf{G}$ is acting. 
If the interaction does not depend on spin variables the strength 
function is thus given by :

\begin{equation}\label{eq:stren}
S(\omega)=-\frac{1}{\pi}Im \int F^{11*}({\bf r}){\bf{
G}}^{11}({\bf r},{\bf r}';\omega)F^{11}({\bf r}')
d{\bf r}~d{\bf r}'
\end{equation}

In the equations above we have not introduced explicitly the
isospin degree of freedom. This can be done directly on the
final equations by doubling the dimension of the matrices in order to
distinguish between neutrons and protons.

\subsection{Spherical symmetry}

In the case of spherical symmetry each qp state is denoted by
the quantum numbers $(E,l,j,m)$, where $E$ is the qp energy and
$(ljm)$ are the standard notations for the orbital and total angular
momenta. Performing the summation over the projection of
the total angular momentum and over the spin variables one gets for
the unperturbed Green function (\ref{eq:g0}) :

\begin{equation}\label{eq:angurad}
{\bf{G}_0}^{\alpha\beta}({\bf r},{\bf r}';\omega)=
\sum_{l_pj_p;l_qj_q \atop LM}
|A^L_{l_pj_p;l_qj_q}|^2
g^{\alpha\beta}_{l_pj_p;l_qj_q}(r,r';\omega)Y_{LM}(\hat r)
Y^*_{LM}(\hat r')
\end{equation}
where the expression for the geometrical coefficients
$A^L_{l_pj_p;l_qj_q}$ is :

\begin{equation}\label{eq:apq}
A^L_{l_pj_p;l_qj_q}=\sqrt{\frac{2j_p+1}{8\pi}}\left(1+(-)^{l_p+l_q+L}\right)
\left(j_p\frac{1}{2}L0\mid j_q\frac{1}{2}\right)
\end{equation}

This expression holds for a residual interaction without gradient terms.
We do not give here the expressions for the case involving gradient terms
since in the calculations of Section III we use a Landau-Migdal form for the
residual interaction.

The radial Green functions $g^{\alpha\beta}_{l_pj_p;l_qj_q}(r, r';\omega)$
are expressed in terms of the qp energies $E_k$ and the
corresponding radial HFB wave functions, i.e.,
$u_k(r) = u_{l_k,j_k}(E_k,r)$ and $v_k(r) = v_{l_k,j_q}(E_k,r)$.
Thus, the radial Green function for a given pair of quantum numbers
$(l_q j_q,l_p j_p)$ is given by:

\begin{equation}\label{eq:g0rad}
g^{\alpha\beta}_{l_pj_p;l_qj_q}(r,r';\omega)=
\sum_{E_p,E_q} \hspace{-0.7cm}\int \hspace{0.5cm} 
\frac{{\cal U}^{\alpha 1}_{pq}(r)
\bar{{\cal U}}^{\beta 1}_{pq}(r')}{\hbar\omega-(E_p+E_q)+i\eta}
-\frac{{\cal U}^{\alpha 2}_{pq}(r)
\bar{{\cal U}}^{\beta 2}_{pq}(r')}{\hbar\omega+(E_p+E_q)+i\eta}
\end{equation}

where
\begin{equation}\label{eq:ubig2}
{\cal U}_{pq}(r)=\left(
        \begin{array}{cc}
          u_p(r)v_q(r) &  u_q(r)v_p(r) \\
        - u_p(r)u_q(r) &  v_p(r)v_q(r)  \\
          v_p(r)v_q(r) &  - u_p(r)u_q(r)   \\
        \end{array}
        \right),\bar{{\cal U}}_{pq}={\cal U}_{pq}+{\cal U}_{qp}
\end{equation}
In Eq. (\ref{eq:g0rad}) the
$\sum \hspace{-0.4cm}\int$ 
~symbol indicates that the
summation is taken both over the discrete and the continuum qp
states.

The unperturbed Green function expressed by Eq. (\ref{eq:angurad})
can be used together with an interaction which does not depend
on spin variables in order to get the radial QRPA Green function,
given by Eq. (\ref{eq:bs}).

All the equations should be written in neutron-proton formalism. Thus, each
supermatrix ($\bf{G}_0$, $\bf{G}$, $\bf{V}$) is divided in 9 blocks
corresponding to the ph, pp, hh case and each one of these blocks is divided
in 4 sub-blocks corresponding to the nn, np, pn and pp quantities. The way to
calculate explicitly the residual interaction supermatrix is given in
appendix.

\subsection {The energy-weighted sum rule in the QRPA}

It is often stated that the Thouless theorem \cite{th61} concerning the
energy-weighted sum rule (EWSR)
of RPA is also valid for the QRPA. 
We give here an explicit proof of this
theorem for the non-trivial case of QRPA.

The Thouless theorem extended to the QRPA means that the equality :

\begin{equation}\label{eq:thouqrpa}
\sum_{\nu}E_\nu\mid \langle\nu\mid F\mid
QRPA\rangle\mid^2=\frac{1}{2}\langle HFB
\mid\left[F,\left[H,F\right]\right]\mid HFB\rangle
\end{equation}
must be satisfied. In the above equation,
$\mid HFB\rangle$ is the HFB ground state and $\mid QRPA\rangle$
is the correlated QRPA ground state while $\mid \nu \rangle$ stands for the
excited QRPA states.

To demonstrate this equality we use the QRPA equations in configuration
space
\cite{rs80}. The specific point of the demonstration 
is that the one-body operator F is now expressed in terms of qp
operators :

\begin{equation}\label{eq:fqrpa}
F=\sum_{kk'}f_{kk'}c^\dagger_k c_{k'}=\sum_{kk' \atop ll'}f_{kk'}
\left[U^*_{kl}U_{k'l'}\beta^\dagger_l\beta_{l'}+
U^*_{kl}V_{k'l'}\beta^\dagger_l\beta^\dagger_{l'}+
V^*_{kl}U_{k'l'}\beta_l\beta_{l'}+
V^*_{kl}V_{k'l'}\beta_l\beta^\dagger_{l'}\right]
\end{equation}

The calculation of the l.h.s.  
of Eq. (\ref{eq:thouqrpa}) 
introduces quantities such as $\langle QRPA \mid F\mid\nu\rangle$. They can
be written as :   

\begin{equation}\label{eq:fxqrpa}
\langle QRPA \mid F\mid\nu\rangle=      \left(\begin{array}{cc}
         \bar{f}^T & \bar{f}^\dagger \\   
        \end{array}\right)
        \left(\begin{array}{c}
         X^\nu \\
         Y^\nu 
        \end{array}\right)
\end{equation}

 where :
\begin{equation}\label{eq:fbarqrpa}
\bar{f}_{ll'}=\sum_{kk'}f_{kk'}h_{kk'll'},
\end{equation}

\begin{equation}\label{eq:hqrpa}
h_{kk'll'}=V^*_{kl'}U_{k'l}-V^*_{kl}U_{k'l'}
\end{equation}

Eq. (\ref{eq:fxqrpa}) allows us to get for  the l.h.s. of Eq.
(\ref{eq:thouqrpa}) an expression similar to the one of
Ref.\cite{th61} (Eq. (36)), i.e. : 

\begin{equation}\label{eq:lhsqrpa}
S_1=    \frac{1}{2} \left(\begin{array}{cc}
         \bar{f}^T & -\bar{f}^\dagger \\
        \end{array}\right)
        \left(\begin{array}{cc}
         A & B \\
         B^* & A*  \\
        \end{array}\right)
        \left(\begin{array}{c}
        \bar{f}^* \\
        -\bar{f}  \\
        \end{array}\right)
\end{equation}

 The r.h.s. of Eq. (\ref{eq:thouqrpa}) is obtained by using Eq.
(\ref{eq:fqrpa}) and the definition of the A and B matrices of QRPA :

\begin{equation}\label{eq:aqrpa}
A_{kk'll'}=\langle HFB
\mid\left[\beta_{k'}\beta_{k},\left[H,\beta^\dagger_{l}\beta^\dagger_{l'}\right]\right]
\mid HFB\rangle,   
\end{equation}

\begin{equation}\label{eq:bqrpa}
B_{kk'll'}=-\langle HFB
\mid\left[\beta_{k'}\beta_{k},\left[H,\beta_{l'}\beta_{l}\right]\right]\mid
HFB\rangle
\end{equation}

with $k<k'$ and $l<l'$. Expressing the r.h.s. with A, B and $\bar{f}$ leads
to Eq. 
(\ref{eq:lhsqrpa}).

This proof is valid for density-independent force. In the case of a 
density-dependent force the relation should remain valid provided the
density-dependent terms commute with F, as it has been shown for the RPA
case \cite{rs80}.

\section{HFB+QRPA calculations of oxygen isotopes}

\subsection{HFB calculations}

We apply our formalism to the calculation of  
neutron-rich oxygen isotopes $^{18,20,22,24}$O.
The ground states are calculated within the continuum HFB approach
\cite{gr01} where the continuum is treated exactly. 
The HFB equations are solved in coordinate space 
with a step of 0.25 fm for the radial coordinate.
In the HFB the mean field quantities are calculated by using the Skyrme
interaction SLy4 \cite{ch98}, while for the pairing interaction 
we take a zero-range density-dependent interaction given by:

\begin{equation}\label{eq:vpair}
V_{pair}=V_0\left[1-\left(\frac{\rho(r)}{\rho_0}\right)^\alpha\right]\delta\left({\bf
r_1}-{\bf r_2}\right)
\end{equation}
where V$_0$, $\rho_0$ and $\alpha$ are the parameters of the force. Due to
its zero-range this force should be used in the HFB calculations with a 
cutoff in qp energy. 
To minimize the number of free parameters, we adapt here the
prescription of Refs. \cite{ga99,be91} which relates the energy cutoff with
the V$_0$ value for the free neutron-neutron system.
To extend this prescription to finite nuclei, we use the relation between 
the energy
$\varepsilon$ of the particle inside nucleus and the energy $\varepsilon_0$
of a free particle:
\begin{equation}\label{eq:cutoff}
\varepsilon(k)=\varepsilon^0(k)\frac{m}{m^*}+U_{HF}
\end{equation}
where m$^*$ is the effective mass, $k$ the momentum and U$_{HF}$ the
Hartree-Fock potential. Since $m^*$ depends on the density we take $m^*/m$=0.7
which is the bulk value for $m^*$. From $\epsilon_{cutoff}$ we can deduce
the qp cutoff energy $E_{cutoff}$.
With this prescription we verified that
the calculated HFB neutron pairing gap $\Delta_n$ remains constant for each
couple ($V_0, E_{cutoff}$). 
In the HFB
calculations we choose a qp cutoff energy equal to 50 MeV. Then, the
prescription of Ref.\cite{ga99} gives V$_0$= -415.73~MeV.fm$^3$. The
parameter $\rho_0$ is set to the usual saturation density, 0.16 fm$^{-3}$.
The value of the parameter $\alpha$ is 
chosen so as to reproduce the trend of the experimental gap. Note that the
calculated gap is defined as the integral of the pairing field whereas the
experimental gap is related to mass differences of the neighboring nuclei
and therefore, there is no need to have an exact quantitative agreement.
We find that the best choice is 
$\alpha$=1.5. Note that the trend of the 
experimental gap is at variance with
the empirical rule $\Delta=12/\sqrt{A}$ MeV.  All the pairing gap values are
displayed in Table \ref{tab:delta}. It should be noted that, in the case of
HFB+Skyrme calculations, if one has a reasonable pairing gap in $^{18,20}$O 
then the pairing is weaker in $^{22}$O and absent in $^{24}$O. This is due
to the 1d$_{5/2}$ and 2s$_{1/2}$ subshell closure. A similar trend is
observed in calculations using Gogny interaction\cite{libert}. 
As seen in Table I, in the
case of QRPA calculations using a Woods-Saxon potential\cite{ma01} 
the gaps are larger 
due to the fact that the energy distance between the relevant 
subshells is smaller.

\subsection{QRPA calculations}

In the QRPA calculations the residual interaction is derived in principle
from the
interaction used in the HFB, i.e., the Skyrme force and the pairing force
(\ref{eq:vpair}).
The zero-range part of the forces pose no problem. The velocity-dependent
terms of the Skyrme force bring additional complications which can be avoided
by approximating the residual interaction in the (ph,ph) subspace by its
Landau-Migdal limit\cite{ba75} where the interacting particle and hole have 
the Fermi momentum and the transferred momentum is zero.
The Skyrme interaction has only $l=0$ and 
$l=1$ Landau parameters. Taking the Landau-Migdal form for the (ph,ph)
interaction simplifies greatly the numerical task, at the cost of the loss
of some consistency. In this work we calculate only natural parity (non
spin-flip) 
excitations and we drop the spin-spin part of the (ph,ph) interaction which
plays a minor role. The Coulomb and spin-orbit residual interactions are
also dropped. 

The QRPA Green function can be evaluated starting with the unperturbed Green
function given by Eq. (\ref{eq:g0rad}). The latter is constructed by using
the solutions of the HFB equations, i.e., the qp energies and the
corresponding wave functions $U$ and $V$. All the qp states are included
until an energy cutoff of 50 MeV, allowing pairs of qp energy until 100 MeV.
In a schematic picture, these pairs are representative of excitations from
the $^{16}$O core, and also excitations of the valence neutrons. The
generalized Bethe-Salpeter equation (\ref{eq:bs}) is solved with a step of
0.5 fm and all radial integrals are carried out up to 22.5 fm. The strength
distribution is calculated until $\omega_{Max}$=50 MeV, with a step of 100
keV and an averaging width $\eta$=150 keV. We have studied two variants of
calculations, the full continuum variant and a box variant where the qp
spectrum is discretized by calculating the HFB solutions with a box boundary
condition, the box radius being 22.5 fm. For $^{24}$O only box calculations
have been performed.

In a fully consistent calculation the spurious center-of-mass state should
come out at zero energy. Because of the Landau-Migdal form of the
interaction adopted here the consistency between mean field and residual qp
interaction is broken and the spurious state becomes imaginary. We cure this
defect by renormalizing the residual interaction by a factor $\alpha$. We
find that in all cases the spurious state $J^{\pi} = 1^{-}$ comes out at
zero energy for $\alpha$=0.80 . 
We have checked that the EWSR are satisfied within 1\% to 5\%.

\subsection{Quadrupole excitations in oxygen isotopes}

We calculate quadrupole strength distributions with the operators
$F_0=\sum_i r_i^2 Y_{20}(\hat{r_i})$ (isoscalar) and
$F_0=\sum_i r_i^2 Y_{20}(\hat{r_i}) t_z(i)$ (isovector). All results
presented correspond to the SLy4 interaction except when stated otherwise.
The strength distributions calculated in
the neutron-rich oxygen isotopes are displayed in Fig.~1.

One can identify a strong low-lying state and the giant quadrupole
resonance (GQR).
The low-lying state becomes more isospin-admixed as the neutron excess
increases.
In the case of $^{24}$O the strength distribution is similar to that
calculated in Ref.\cite{ma01} with a Woods-Saxon potential for the mean field
although pairing effects are negligible in our calculation whereas the gap
$\Delta$ of Ref.\cite{ma01} is sizable. The main difference is the
position of the first 2$^+$ state located at 4.0 MeV here and  5.0
MeV in Ref.\cite{ma01}. In the other nuclei this low-lying state is at lower
energies.
This is due to the 2s$_{1/2}$ subshell
closure in $^{24}$O. The HF single-particle energies are given in Table
\ref{tab:hfox}. The 2s$_{1/2}$ state is more bound in the $^{24}$O nucleus,
suggesting a stronger subshell closure in this case. The occupation factors
of these states calculated in HFB are displayed in Table
\ref{tab:hfox}. The 2s$_{1/2}$ starts to be significantly populated in $^{22}$O
due to the pairing correlations. In the $^{18,20}$O spectra mainly 3
low-lying peaks are present. In the unperturbed case they correspond to the
(1d$_{5/2}$,1d$_{5/2}$), (1d$_{5/2}$,2s$_{1/2}$) and (1d$_{5/2}$,1d$_{3/2}$)
two-qp neutron configurations. Their energies are given in Table
\ref{tab:unper}. In $^{18,20}$O the configuration (1d$_{5/2}$,1d$_{3/2}$)
has a very low strength whereas the (1d$_{5/2}$,1d$_{5/2}$),
(1d$_{5/2}$,2s$_{1/2}$) configurations have similar strength. The effect of
the residual interaction, in addition to admix the configurations, is to
lower the energy of the initial (1d$_{5/2}$,1d$_{5/2}$) peak and to increase
the strength of the low-lying state (cf. Fig. \ref{fig:oxystrength}).

The effect of the
residual interaction in $^{22}$O is displayed in Fig.\ref{fig:vres}, showing
the isoscalar strength functions calculated with the unperturbed Green
function G$_0$ and with the full G. The effect of the residual
interaction is to gather the strength to generate collective modes such as
the GQR and the low-lying state.

The $E2$ energy and $B(E2)$ value
of the first 2$^+$ state are displayed in Table
\ref{tab:be2}. As noted before, the $E2$ energy in $^{24}$O is larger than in
other isotopes due to the 2s$_{1/2}$ subshell closure. The $E2$ energy in
$^{18}$O is overestimated and that of $^{22}$O is underestimated as compared
to experiment. This shows, as noted previously in QRPA calculations with a
constant gap (\cite{kh00}), that the energy prediction of the low-lying
modes is a delicate task in RPA-type models. The $B(E2)$ values are well
reproduced except for the problematic $^{18}$O nucleus. This discrepancy
observed for the $B(E2)$ value of $^{18}$O is also found in several
shell-model calculations\cite{br88,ut99} and in previous QRPA
calculations\cite{kh00}, showing a limitation of such models to study
$^{18}O$ low-lying states. Indeed this high B(E2) anomaly may be due to
the presence of deformed states in the experimental low-lying spectra of
$^{18}O$ \cite{br66}. Moreover the observation of many low-lying states in
this nucleus can be described by bands characteristics of the rotational
spectra. It has been suggested that the low-lying states in these nuclei
(such as $^{16,17,18}O$) can be described as a mixture between highly
deformed states and the usual shell model states \cite{br66,fe65}. This
allows to successfully reproduce both the E2 and B(E2) of the low-lying
states. As stated in ref.
\cite{br66}, for heavier oxygen isotopes, the energies of the deformed states
become higher, and thus the admixture is smaller. This may explain why the
calculated $B(E2)$ are in good agreement with the experimental data for
$^{20,22}$O nuclei. The $B(E2)$ in $^{24}$O is predicted smaller than those
of lighter isotopes, which supports the 2s$_{1/2}$ subshell closure effect.
In order to display the structure of the low-lying sector, the calculated
energies and B(E2) of the second and third 2$^+$ states are shown in Table
\ref{tab:second} for the oxygen isotopes. The 2$^+_2$ and 2$^+_3$ energies
are overestimated in the case of $^{18}O$, whereas a good agreement is found
for the energy of the 2$^+_2$ state for $^{20}O$. This may support the presence
of deformed admixtures in the light neutron-rich oxygen isotopes such as
$^{18}O$.

The calculated M$_n$/M$_p$ ratios indicate that the neutron are more
coherently contributing to the excitation when their number is increasing.
For example the M$_n$/M$_p$ ratio for $^{24}$O is more than twice the N/Z
value, indicating a very strong neutron contribution to the excitation. The
calculated M$_n$/M$_p$ ratio is correctly reproducing that of $^{20}$O
deduced from proton scattering experiments. In the case of $^{18}$O the
experimental $M_n/M_p$ is not well reproduced. This is linked to the fact
that the $B(E2)$ value is not well described by the model. The transition
densities for the first low-lying 2$^+$ state of the neutron-rich oxygen
isotopes are displayed on figure
\ref{fig:trans}. In the case of $^{22,24}$O the neutron transition density
is located more on the surface than the proton one, possibly indicating the
presence of a neutron skin.

The pairing effects are depicted in Fig. \ref{fig:rpa} where the continuum
QRPA calculation is compared to a HF+RPA calculation for the $^{22}$O
nucleus. The effect of pairing is to shift to higher energy the
low-lying peak, and to split the second 2$^+$ state into two states with
smaller strength. There is also some effect in the GQR region. 

In order to investigate the effect of the density
dependence of the pairing interaction we have also calculated the strength
distributions 
with a density-independent interaction, i.e., $\rho_0$
going to infinity in Eq. (\ref{eq:vpair}). In the HFB calculation, the V$_0$
parameter has been chosen to reproduce the experimental gap of $^{18}$O, 
V$_0$=-220 MeV.fm$^3$ (in this case the prescription of Ref.\cite{ga99} is
no longer applied). 
Fig.\ref{fig:densdep} compares the results in $^{18}$O
calculated with the density-dependent and density-independent interactions. 
The effect of the density dependence is to increase the energies of the 2$^+$ 
states, and to slightly lower the strength of the low-lying states.

Box discretization calculations have also been performed in order to test
the box boundary condition approximation. The results are shown in 
Fig.\ref{fig:disc} for $^{22}$O. One can see that
only the low-lying state is nearly insensitive whereas the structure of the 
GQR is more affected by the way the continuum is treated. This shows the
necessity of the exact continuum treatment in order to study the giant
resonances in neutron-rich oxygen isotopes.

Since our full calculations (HFB+QRPA) have only the Skyrme and the pairing
interaction as inputs the results may be used to learn about the different
Skyrme parameterizations. Figure 5 shows a comparison of results obtained
with the SLy4 \cite{ch98}, SGII \cite{gi81} and SIII \cite{bei75} for the
$^{20}$O nucleus. There is no drastic effect depending on the force. The
SIII interaction shifts some low-lying states to higher energy and increases 
the strength of the state located around 5 MeV. All three interactions
produce a splitting of the GQR but 
the SGII force predicts more
strength in the lower component of the giant resonance.

\section{Conclusions}

        We have derived the Bethe-Salpeter equation for the QRPA from the
	small amplitude limit of the perturbed time dependent HFB equations.
	This approach ensures the self-consistency at the conceptual level
	between the mean field, the pairing field and the qp residual
	interaction. The QRPA Green function is decomposed into the ph, pp
	and hh channels.
        The supermatrix representing the residual interaction is determined
	self-consistently from the Skyrme and the pairing interactions used
	in the HFB calculations. The Thouless theorem concerning the EWSR
	sum rule is shown to hold in the case of self-consistent QRPA.

        As an application we have studied the quadrupole excitations of the
	neutron-rich oxygen isotopes using Skyrme-type interactions for the
	mean field and a zero-range, density-dependent interaction for the
	pairing field. In the numerical study we have approximated the ph
	residual interaction coming from the Skyrme force by its Landau
	limit.
        The coupling to the continuum appears to have a sizable effect on
	the GQR and a minor effect on the low-lying states. This shows the
	importance of the full continuum treatment in order to study giant
	resonances in neutron-rich nuclei.
	The low-lying states are sensitive to the pairing interaction. The
	first 2$^+$ state of $^{18}$O is not well described, as previously
	noted with other models such as shell-model calculations. Additional
	investigations on this nucleus are called for. The continuum QRPA
	shows its ability to reproduce the experimental data of the first
	2$^+$ state for heavier oxygen isotopes, and it predicts a lowering
	of the $B(E2)$ value for the $^{24}$O nucleus. A future improvement
	of the model will be to include fully the velocity-dependent terms
	of the Skyrme interaction in the ph channel. Work along these lines
	is in progress.

\appendix
\section{hamiltonian perturbation and residual interaction}
The HFB energy functional is written as follow:
\begin{equation}\label{eq:efunc}
{\cal E}=\sum_{ml}t_{ml}\rho_{ml}+\frac{1}{2}\sum_{mlpq}\langle lq\mid
\bar{{\cal V}}\mid mp\rangle\rho_{pq}\rho_{ml} + \frac{1}{4}\sum_{mlpq}\langle
lm\mid\bar{{\cal V}}_P\mid pq\rangle\kappa^*_{lm}\kappa_{pq}
\end{equation}
with $\bar{{\cal V}}$ the antisymmetrised interaction and $\bar{{\cal V}}_p$
the antisymmetrised pairing interaction.

The HFB hamiltonian is :

\begin{equation}\label{eq:h0}
	{\cal H}^0_{ij}=\left(
	\begin{array}{cc}
	h_{ij} & \Delta_{ij} \\
	\Delta^{\dagger}_{ij} & -h^*_{ji} \\
	\end{array}
	\right)
\end{equation}

with 
\begin{equation}\label{eq:funcder}
h_{ij}=\frac{\partial{\cal E}}{\partial\rho_{ij}},
\Delta_{ij}=\frac{\partial{\cal E}}{\partial\kappa^*_{ij}}
\end{equation}

Next we expand the perturbation of the hamiltonian on the densities
perturbations and get :

\begin{equation}\label{eq:h'1}
	{\cal H}'^{11}_{ij}=	
	\sum_{kl}\frac{\partial^2{\cal
	E}}{\partial\rho_{kl}\partial\rho^*_{ij}}\rho'_{kl}+\frac{1}{2}\frac{\partial^2{\cal
	E}}{\partial\kappa_{kl}\partial\rho^*_{ij}}\kappa'_{kl}+\frac{1}{2}\frac{\partial^2{\cal
	E}}{\partial\bar{\kappa}_{kl}\partial\rho^*_{ij}}\bar{\kappa}_{kl} \\
\end{equation}
\begin{equation}\label{eq:h'2}
	{\cal H}'^{12}_{ij}=
	\sum_{kl}\frac{\partial^2{\cal
	E}}{\partial\rho_{kl}\partial\kappa^*_{ij}}\rho'_{kl}+\frac{1}{2}\frac{\partial^2{\cal
	E}}{\partial\kappa_{kl}\partial\kappa^*_{ij}}\kappa'_{kl}	\\
\end{equation}
\begin{equation}\label{eq:h'3}
        {\cal H}'^{21}_{ij}=
	\sum_{kl}\frac{\partial^2{\cal
	E}}{\partial\rho_{kl}\partial\bar{\kappa}^*_{ij}}\rho'_{kl}+\frac{1}{2}\frac{\partial^2{\cal
	E}}{\partial\bar{\kappa_{kl}}\partial\bar{\kappa}^*_{ij}}\bar{\kappa}'_{kl}\\	
\end{equation}
\begin{equation}\label{eq:h'4}
	{\cal H}'^{22}_{ij}=-{\cal H}'^{11}_{ji} \\
\end{equation}

To get ${\cal H}'$ in coordinate space, we use the relationship between the
densities :
\begin{equation}\label{eq:rhospco}
        \rho\left({\bf r}\sigma\right) = \sum_{ij} \phi^*_i\left({\bf
	r}\sigma\right)\phi_j\left({\bf r}\sigma\right) \rho_{ji}
\end{equation}

\begin{equation}\label{eq:rhotildspco}
        \kappa\left({\bf r}\sigma\right) = \sum_{ij} \phi_i\left({\bf
	r}\bar{\sigma}\right)\phi_j\left({\bf r}\sigma\right) \kappa_{ji}
\end{equation}

\begin{equation}\label{eq:rhotildbarspco}
        \bar{\kappa}\left({\bf r}\sigma\right) = \sum_{ij}
	\phi^*_i\left({\bf
	r}\bar{\sigma}\right)\phi^*_j\left({\bf r}\sigma\right) \kappa^*_{ji}
\end{equation}
where $\phi_i\left({\bf r}\sigma\right)$ is the nucleon wave function.

Using Eq. (\ref{eq:h'1}-\ref{eq:h'4}) together with Eq.
(\ref{eq:rhospco}-\ref{eq:rhotildbarspco}) allows to
calculate ${\cal H}'$ in coordinate space, and get Eq. (\ref{eq:hvar}).

\newpage

\begin{table}[h]
\begin{tabular}{|c|c|c|c|c|}
      & $\Delta_{Exp}$ (MeV) & $\Delta_{Phen}$ (MeV) &
  $\Delta_{HFB}$ (MeV) &   $\Delta_{WS}$ (MeV)
\\ \hline
 $^{18}O$ & 1.95  & 2.83  &  1.96  & 2.74  
\\ 
 $^{20}O$ & 1.83  & 2.68  &  1.85  & 3.13
\\
 $^{22}O$ & 1.52  & 2.56 & 1.04    & 3.30
\\
 $^{24}O$ & 0.49 & 2.45  &  0.00   &3.39
\end{tabular}

\caption{\label{tab:delta}}
Neutron pairing gaps : $\Delta_{Exp}$ is the experimental value taken as the
odd-even mass difference \cite{au95}, $\Delta_{Phen}$ is using the empirical
12/$\sqrt{A}$ MeV prescription \cite{bo69}, $\Delta_{HFB}$ is calculated in the
present work, and $\Delta_{WS}$ are the gap used in ref \cite{ma01}.
\end{table}

\begin{table}[h]
\begin{tabular}{|c||c|c||c|c||c|c||c|c||}
  & \multicolumn{2}{c||}{\textbf{$^{18}O$}} &
  \multicolumn{2}{c||}{\textbf{$^{20}O$}} &
  \multicolumn{2}{c||}{\textbf{$^{22}O$}} &
  \multicolumn{2}{c||}{\textbf{$^{24}O$}}
\\ \hline
 1d$_{5/2}$ & -6.7 & 0.31 & -6.9 & 0.62  & -7.2 & 0.93  & -7.7 & 1.   
\\ \hline
 2s$_{1/2}$ & -4.0 & 0.03 & -4.2 & 0.08  & -4.6 & 0.18 & -4.9 & 1.   
\\ \hline
 1d$_{3/2}$ & 0.3 & 0.01 & 0.3 & 0.02  & 0.2  & 0.01 & 0.2 &0.  
\end{tabular}

\caption{\label{tab:hfox}}
1d$_{5/2}$, 2s$_{1/2}$ and 1d$_{3/2}$ levels in the $^{18,20,22,24}$O
nuclei. For each nucleus the left column shows the single-particle energies
(MeV) calculated with the HF approximation, and the right column displays
the occupation factors for the single-qp levels calculated with the HFB
model.
\end{table}

\begin{table}[h]
\begin{tabular}{|c||c|c|c|}
  & \textbf{$^{18}O$}   & \textbf{$^{20}O$}  & \textbf{$^{22}O$} 
\\ \hline
 (1d$_{5/2}$,1d$_{5/2}$) & 4.52 & 4.16 & 4.60 
\\ \hline
 (1d$_{5/2}$,2s$_{1/2}$) & 5.72 &  4.36 & 3.35    
\\ \hline
 (1d$_{5/2}$,1d$_{3/2}$) & 10.39  & 9.09  & 7.70    
\end{tabular}

\caption{\label{tab:unper}}
Two qp energies (MeV) of the (1d$_{5/2}$,1d$_{5/2}$),
(1d$_{5/2}$,2s$_{1/2}$), and (1d$_{5/2}$,1d$_{3/2}$) configurations of the
unperturbed strength function for the $^{18,20,22}$O nuclei.
\end{table}

\begin{table}[h]
\begin{tabular}{|c||c|c|c|c|}
  & \textbf{$^{18}O$}   & \textbf{$^{20}O$}  & \textbf{$^{22}O$} &
  \textbf{$^{24}O$} 
\\ \hline
 $E2$ (MeV) & 3.2 /(2.0$^{a)}$) & 2.3 /(1.7$^{a)}$) & 1.9 /(3.2$^{b)}$) & 4.0   
\\ \hline
 B(E2) $(e^2fm^4)$& 14  /(45 $\pm 2^{c)}$) & 22 / (28 $\pm 2^{c)}$) & 22/
 (21 $\pm 8^{d)}$) & 9  
\\ \hline
$(M_n/M_p)_{2^+}$& 2.88  /(1.10 $\pm$0.24$^{e)}$)& 3.36 / (3.25
$\pm$0.80$^{e)}$) & 3.53 & 4.37
\end{tabular}

\caption{\label{tab:be2}}
Energy, proton contribution to the reduced transition probabilities B(E2),
and ratio of the transition matrix elements $M_n/M_p$ for the first 2$^+$
state in the $^{18,20,22,24}$O nuclei, calculated with the present model.
Measured E2, B(E2) values and the $M_n/M_p$ ratios corresponding to the
experimental data are displayed in brackets.\\
a) ref. \cite{fi96} ;
b) ref. \cite{be99} ;
c) ref. \cite{ra87} ;
d) ref. \cite{th00} ;
e) ref. \cite{kh00}
\end{table}

\begin{table}[h]
\begin{tabular}{|c||c|c|c|c|}
  & \textbf{$^{18}O$}   & \textbf{$^{20}O$}  & \textbf{$^{22}O$} &
  \textbf{$^{24}O$} 
\\ \hline
\hline
2$^+_2$ : E2 (MeV) & 5.3  /(4.0$^{a)}$)& 4.2  /(4.1$^{a)}$)& 6.2  & 7.1   
\\ \hline
2$^+_2$ : B(E2) $(e^2fm^4)$& 1.0 & 0.3  & 1.4  & 4.0  
\\ \hline
\hline
2$^+_3$ : E2 (MeV) & 9.8 /(5.3$^{a)}$) & 8.3 /(5.2$^{a)}$) & 7.5  & 8.1   
\\ \hline
2$^+_3$ : B(E2) $(e^2fm^4)$& 1.5   & 2.5  & 2.5  & 0.7  
\end{tabular}

\caption{\label{tab:second}}
Energy and proton contribution to the reduced transition probabilities B(E2)
in the $^{18,20,22,24}$O nuclei, calculated with the present model for the
2$^+_2$ (upper lines) and 2$^+_3$ (lower lines) states. Measured E2 values
corresponding to the experimental data are displayed in brackets.
\\ 
a) ref.\cite{fi96} ;
\end{table}

\begin{figure}[h]
\vspace{0.0cm}
\caption {Isoscalar (solid) and isovector (dashed) quadrupole 
strength functions calculated in continuum-QRPA for the
$^{18,20,22,24}$O isotopes.}
\label{fig:oxystrength}
\end{figure}

\begin{figure}[h]
\vspace{0.0cm}
\caption {Isoscalar quadrupole strength function 
  calculated in continuum-QRPA for the $^{22}$O
nucleus. The unperturbed strength (dashed line) is also shown.} 
\label{fig:vres}
\end{figure}

\begin{figure}[h]
\vspace{0.0cm}
\caption {Neutron and proton transition densities of the first 2$^+$ state of 
$^{18,20,22,24}$O nuclei.}
\label{fig:trans}
\end{figure}

\begin{figure}[h]
\vspace{0.0cm}
\caption {Isoscalar strength function calculated in continuum-QRPA
(solid line) and HF+RPA (dashed line) with box boundary conditions for the $^{22}$O
nucleus.}
\label{fig:rpa}
\end{figure}

\begin{figure}[h]
\vspace{0.0cm}
\caption {Isoscalar strength function calculated with a density-independent 
pairing interaction (solid line) and density-dependent pairing interaction
(dashed line) with box boundary conditions for the $^{18}$O nucleus.}
\label{fig:densdep}
\end{figure}

\begin{figure}[h]
\vspace{0.0cm}
\caption {Isoscalar strength function calculated in continuum-QRPA
(solid line) and with a box discretization (dashed line) 
for the $^{22}$O nucleus.}
\label{fig:disc}
\end{figure}

\begin{figure}[h]
\vspace{0.0cm}
\caption {Isoscalar quadrupole
strength function calculated in continuum-QRPA for the $^{20}$O nucleus with
various Skyrme interactions : SLy4 (solid line), SGII (dashed line) and SIII
(dotted line).}
\label{fig:skyrme}
\end{figure}




\begin{references}
\bibitem{rs80} P. Ring, P. Schuck, {\it The nuclear many-body problem,
Springer-Verlag} (1980).
\bibitem{av60} R. Arvieu and M. V\'en\'eroni, Compt. Rend. Acad. Sci. {\bf
    250} (1960) 992, 2155.
\bibitem{km60} M. Kobayasi and T. Marumori, {\it Prog. Theor. Phys.} {\bf 23}
  (1960) 387.  
\bibitem{ba60} M. Baranger, {\it Phys. Rev.} {\bf 120} (1960) 957.
\bibitem{be75} G.F.Bertsch and S. F. Tsai, {\it Phys. Rep.} {\bf 18} (1975) 125.
\bibitem{lg76} K.F. Liu and Nguyen Van Giai, {\it Phys. Lett.} {\bf B 65} (1976) 23.  
\bibitem{gi81} Nguyen Van Giai and H. Sagawa, {\it Nucl. Phys} {\bf A371}
(1981) 1.
\bibitem{ka98} S. Kamerdzhiev, R. J. Liotta, E. Litvinova, and
V. Tselyaev,  Phys. Rev {\bf C58} (1998) 172.
\bibitem{ha82} K. Hagino and H. Sagawa, {\it Nucl. Phys} {\bf A695}
(2001) 82.
\bibitem{ma01} M. Matsuo, {\it Nucl. Phys.} {\bf A696} (2001) 371.
\bibitem{gr01} M. Grasso, N. Sandulescu, Nguyen Van Giai, R. J. Liotta, 
{\it Phys. Rev} {\bf C64} (2001) 064321.
\bibitem{th61} D. J. Thouless , {\it Nucl. Phys} {\bf 22} (1961) 78.
\bibitem{ch98} E. Chabanat, P. Bonche, P. Haensel, J. Meyer, R. Schaeffer, 
{\it Nucl. Phys.} {\bf A635} (1998) 231.
\bibitem{ga99} E. Garrido, P. Sarriguren, E. Moya de Guerra, P. Schuck, 
{\it Phys. Rev} {\bf C60} (1999) 064312.
\bibitem{be91} G. F. Bertsch, H. Esbensen, {\it Ann. of Phys.} {\bf 209}
(1991) 327.
\bibitem{libert}J. Libert, private communication.
\bibitem{ba75} S. O. B\"ackman, A. D. Jackson and J. Speth, 
{\it Phys. Lett.} {\bf B56} (1975) 209.
\bibitem{kh00} E. Khan, Y. Blumenfeld, Nguyen Van Giai, T. Suomijarvi, N. Alamanos,
F. Auger, G. Colo, N. Frascaria, A. Gillibert, T. Glasmacher, M. Godwin,
K. W. Kemper, V. Lapoux, I. Lhenry, F. Marechal, D. J. Morrissey, A. Musumarra,
N. A. Orr, S. Ottini-Hustache, P. Piattelli, E. C. Pollacco, P. Roussel-Chomaz,
J. C. Roynette, D. Santonocito, J. E. Sauvestre, J. A. Scarpaci, C. Volpe, {\it
Phys. Lett.} {\bf B490} (2000) 45.
\bibitem{br88} B. A. Brown and B. H. Wildenthal, {\it Ann. Rev. Part. Nucl.
Sci.} {\bf 38} (1988) 29.
\bibitem{ut99} Y. Utsuno, T. Otsuka, T. Mizusaki, M. Honma,
{\it Phys. Rev} {\bf C60} (1999) 054315.
\bibitem{br66} G. E. Brown and A. M. Green, {\it Nucl. Phys.} {\bf 75} (1966)
401.
\bibitem{fe65} P. Federman and I. Talmi, {\it Phys. Lett.} {\bf 15} (1965)
165.
\bibitem{bei75} M. Beiner, H. Flocard, Nguyen Van Giai, P. Quentin, 
{\it Nucl. Phys} {\bf A238} (1975) 29.
\bibitem{au95} G. Audi and Wapstra, {\it Nucl. Phys} {\bf A595} (1995) 409.
\bibitem{bo69} A. Bohr and B. Mottelson,  {\it Nuclear Structure} (Benjamin,
New York, 1969) Vol. 1, p. 169.
\bibitem{fi96}  R. B. Firestone, {\it Table of isotopes} Eighth edition (1996).
\bibitem{be99}  M. Belleguic, M. J. Lopez-Jimenez, M. Stanoiu, F. Azaiez,
M.-G. Saint-Laurent, O. Sorlin, N. L. Achouri, J.-C. Angelique, C. Bourgeois,
C. Borcea, J.-M. Daugas, C. Donzaud, F. De Oliveira-Santos, J. Duprat, S. Grevy,
D. Guillemaud-Mueller, S. Leenhardt, M. Lewitowicz, Yu.-E. Penionzhkevich,
Yu. Sobolev, {\it Nucl. Phys} {\bf A682} (2001) 136c.
\bibitem{ra87} S. Raman, C. H. Malarkey, W. T. Milner, C. W. Nestor Jr. , 
P. H. Stelson, {\it Atomic Data and Nuclear Data Tables } {\bf 36} (1987) 1.
\bibitem{th00} P. G. Thirolf, B. V. Pritychenko, B. A. Brown, P. D. Cottle,
M. Chromik, T. Glasmacher, G. Hackman, R. W. Ibbotson, K. W. Kemper, T.
Otsuka, L. A. Riley, H. Scheit, {\it Phys. Lett.} {\bf B485} (2000) 16.
\end{references}
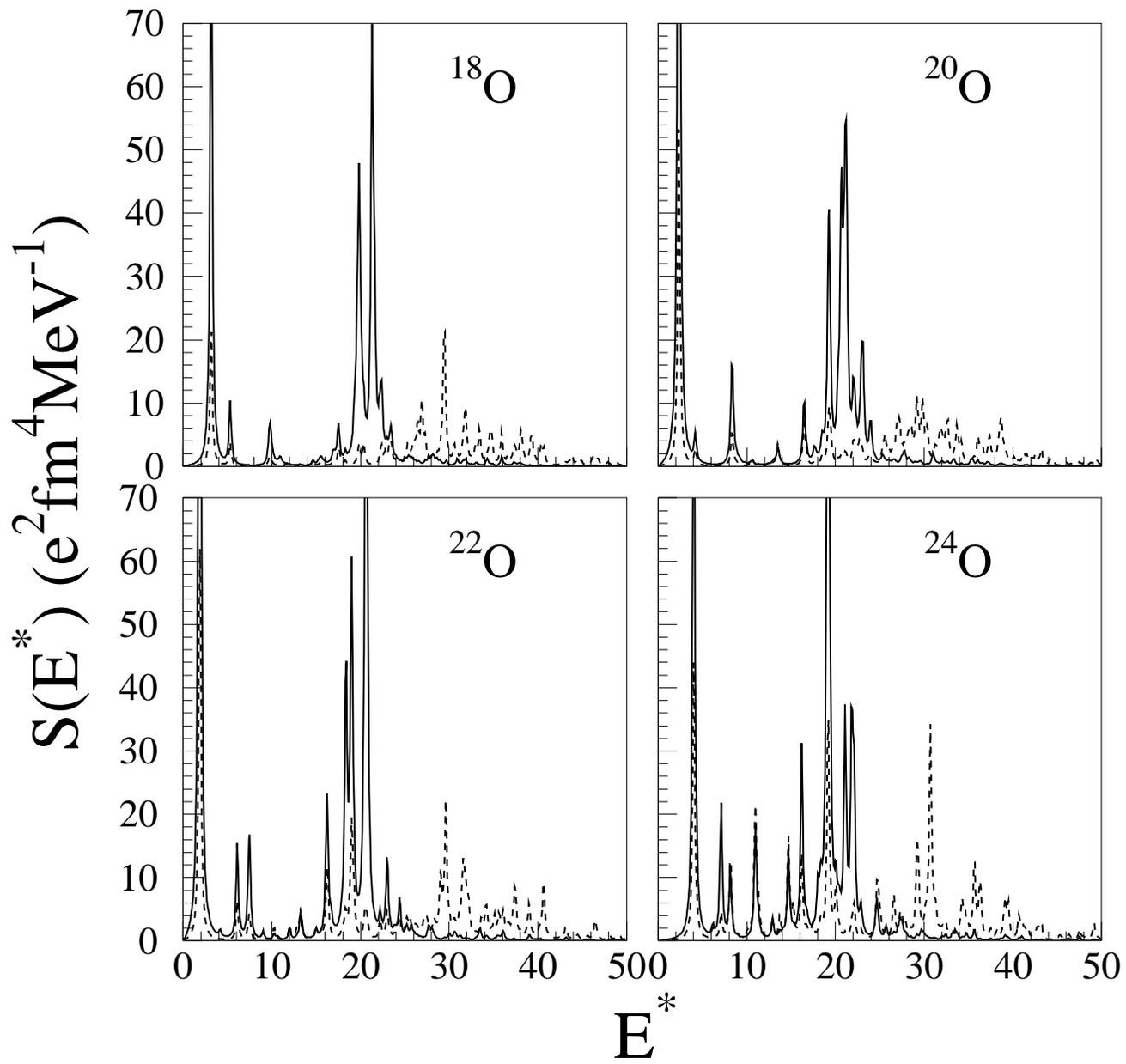
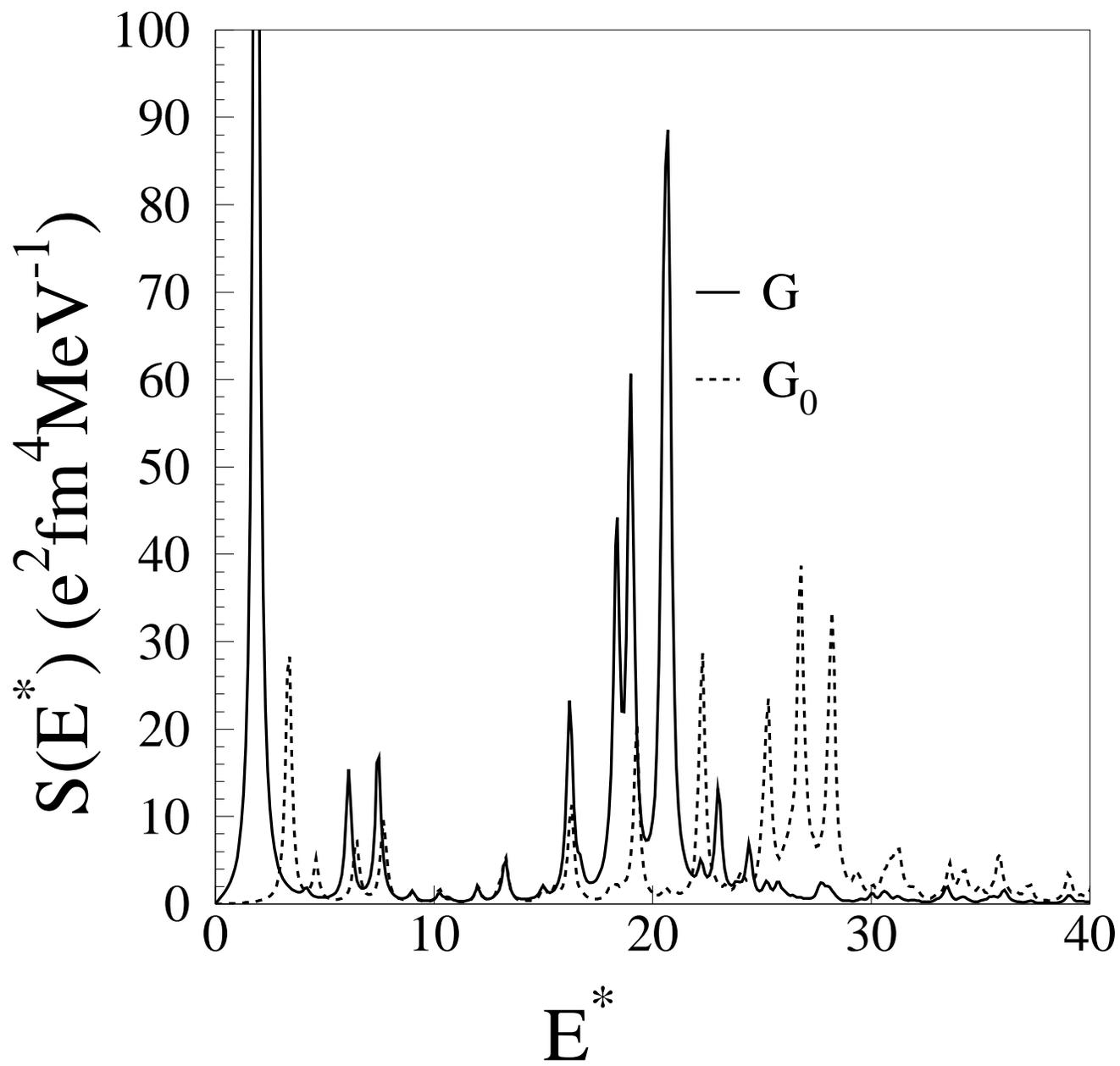
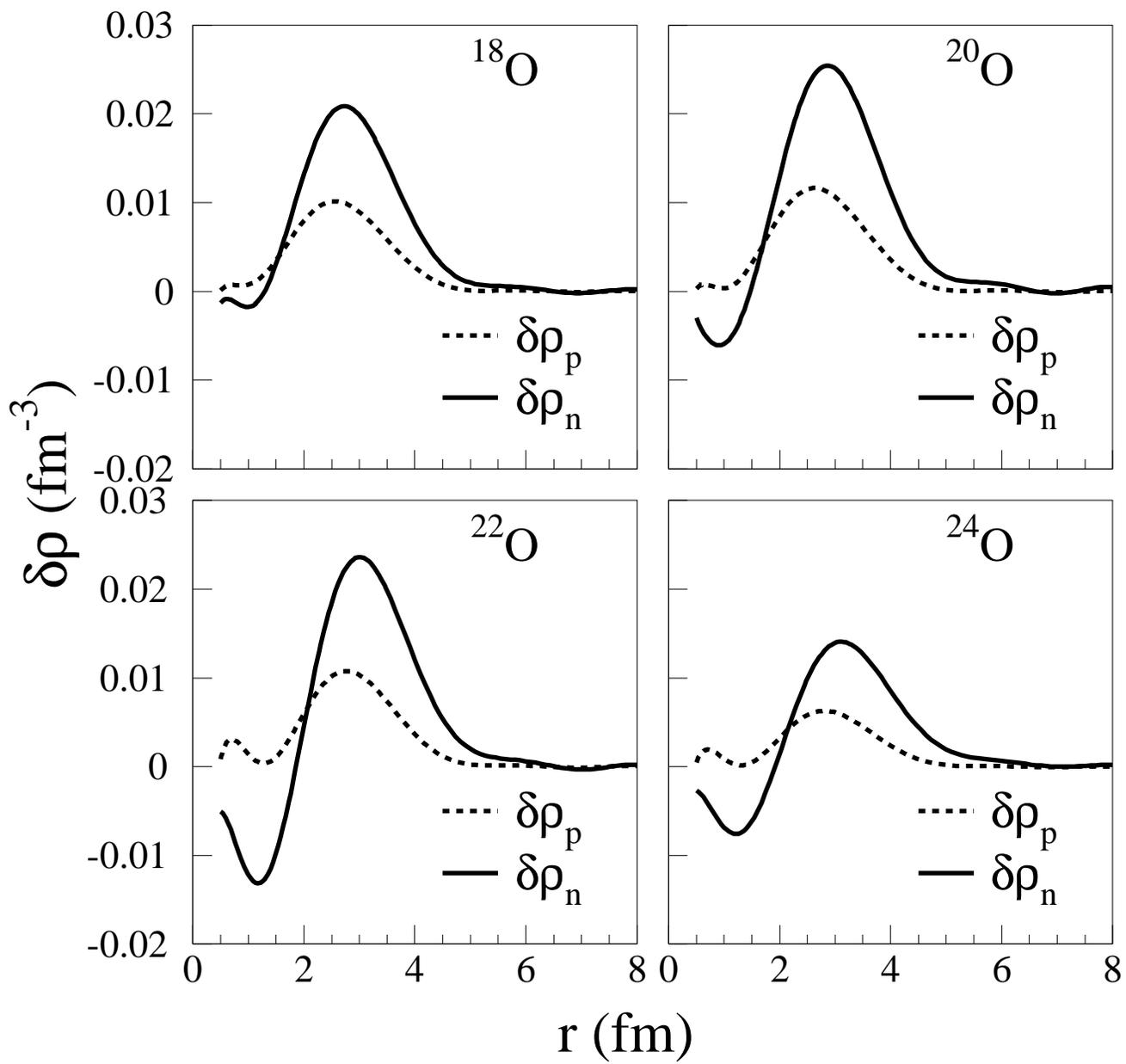
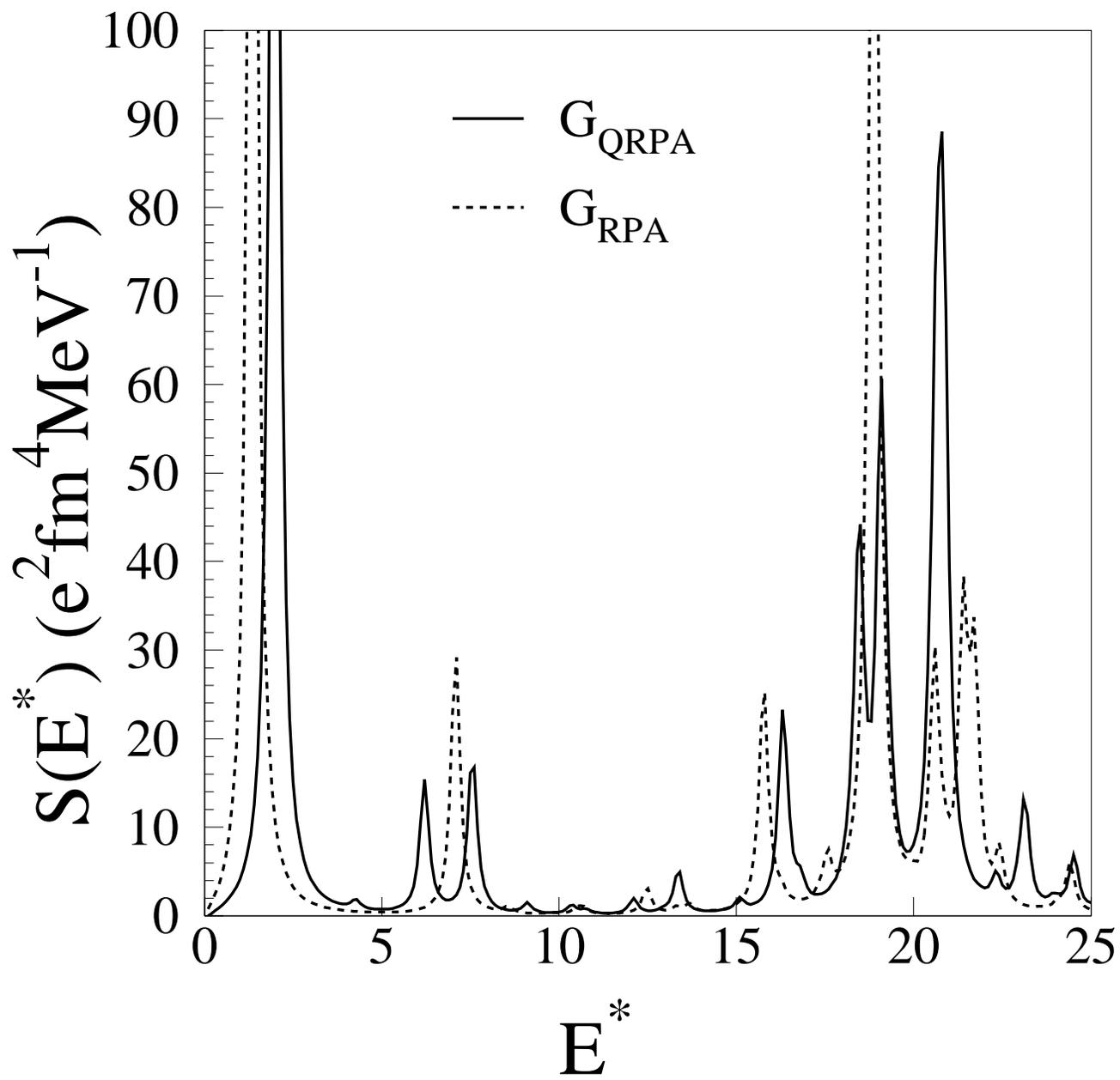
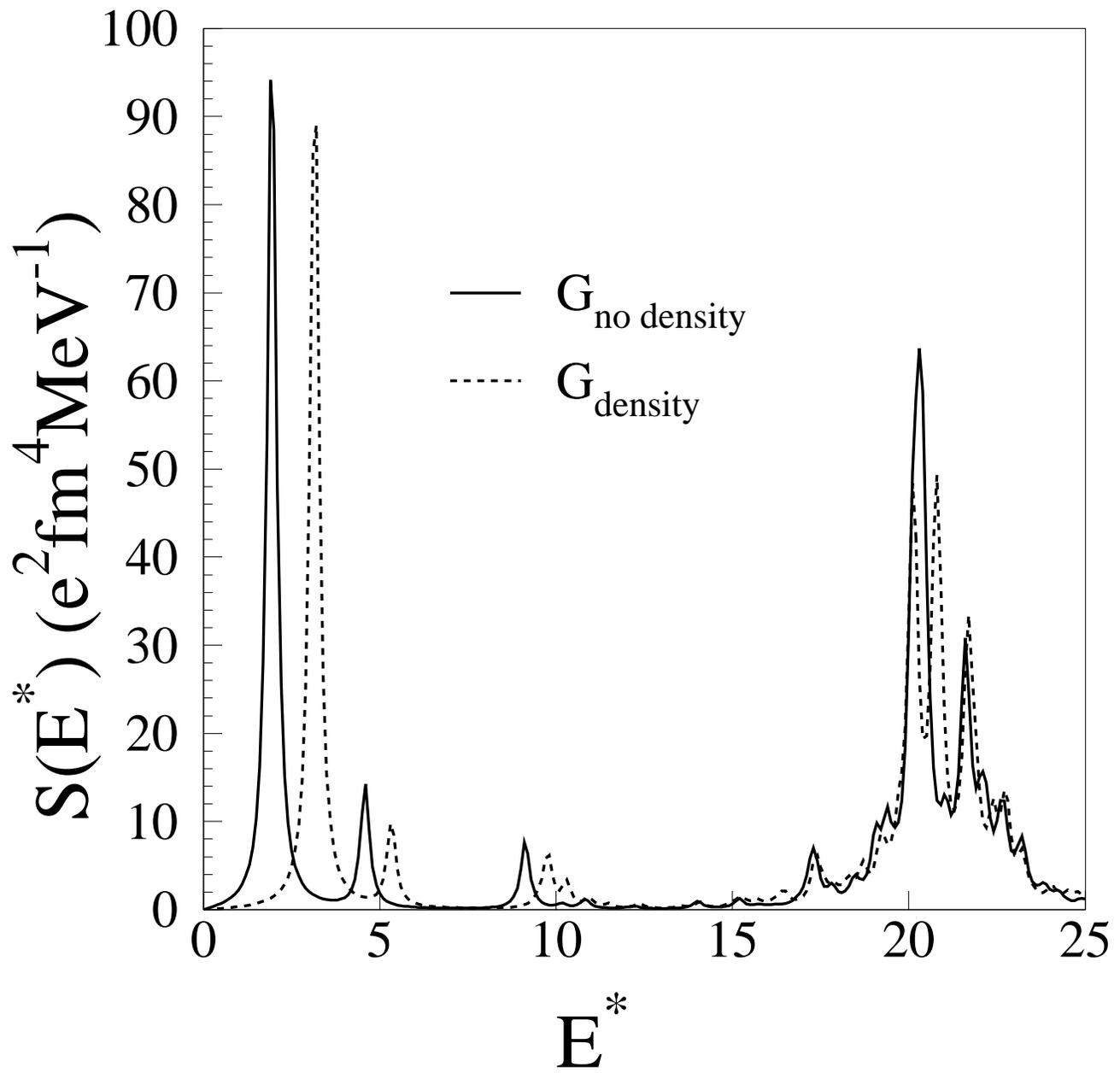
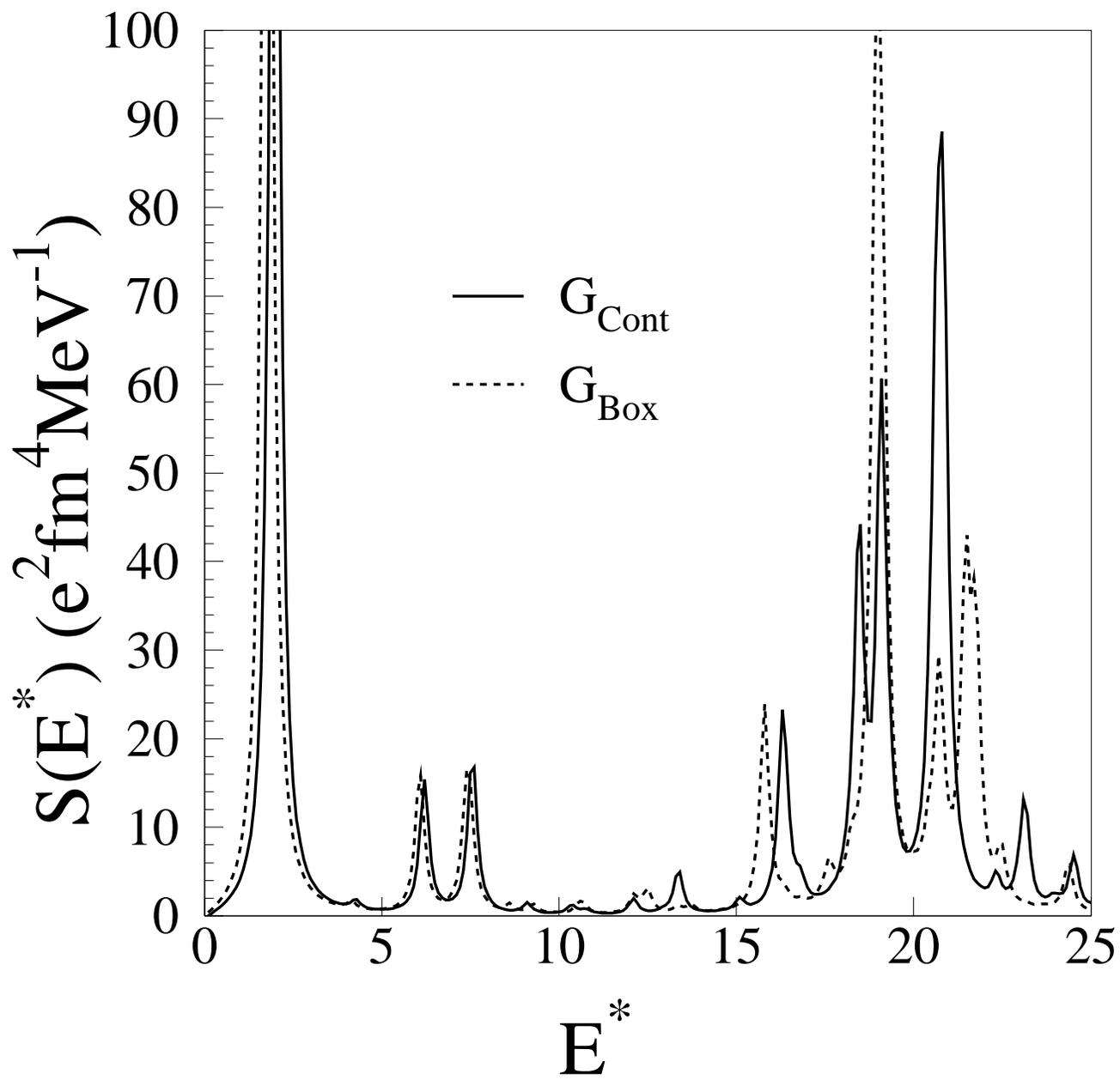
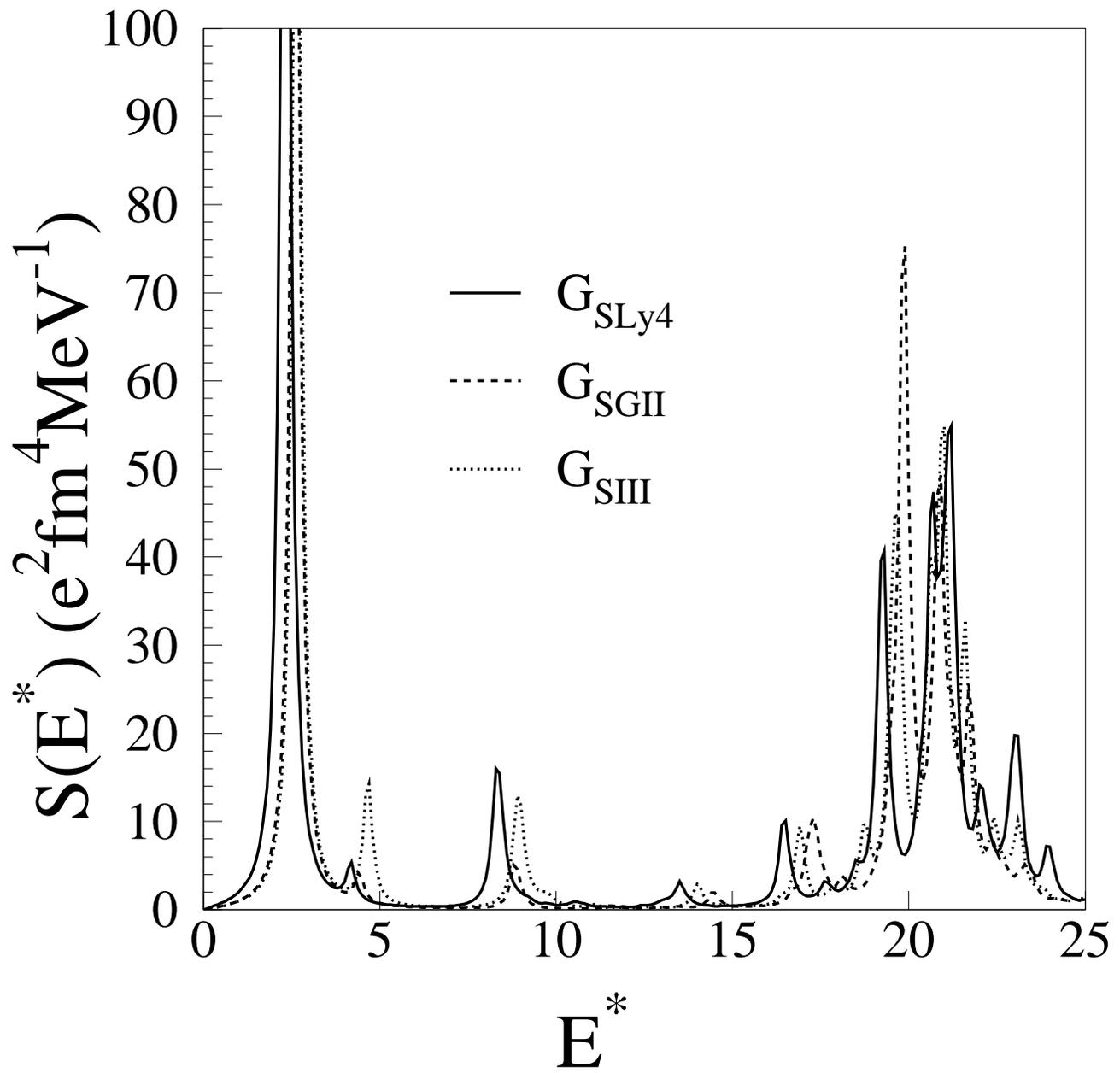
\end{document}